\documentclass[preprint,showpacs,preprintnumbers,amsmath,amssymb]{revtex4}

\usepackage{graphicx}
\usepackage{dcolumn}
\usepackage{bm}
\usepackage{epsfig}

\newcommand{\be}{\begin{eqnarray}}
\newcommand{\ee}{\end{eqnarray}}
\def\refeq#1{(\ref{#1})}

\def\nn{\nonumber}

\def\L{\Lambda}

\def\l{\left}
\def\r{\right}
\def\te{\mbox{e}}
\def\rmi{\mbox{i}}
\def\down{\downarrow}
\def\up{\uparrow}

\begin{document}

\title{Exact results for the $1$D interacting mixed Bose-Fermi gas}

\author{M.T. Batchelor, M. Bortz, X.W. Guan, N. Oelkers}

\affiliation{Department of Theoretical Physics, Research School of Physical Sciences and Engineering\\
and Department of Mathematics, Mathematical Sciences Institute,\\
Australian National University, Canberra ACT 0200,  Australia}

\date{\today}

\begin{abstract}
\noindent
The exact solution of the $1$D interacting mixed Bose-Fermi gas is used to calculate ground-state properties both for finite systems and in the thermodynamic limit.
The quasimomentum distribution, ground-state energy and generalized velocities are obtained as functions of the interaction strength both for polarized and non-polarized fermions.
We do not observe any demixing instability of the system for repulsive interactions.
\end{abstract}

\pacs{03.75.Ss, 05.30.Fk, 67.60.-g,71.10.Pm}

\keywords{Quantum mixed Fermi gas, momentum distribution function, phase separation}

\maketitle

The cooling and trapping of quantum Fermi gases of ultracold atoms poses 
a number of additional challenges to those faced for bosons \cite{BEC}.
A key point is that no more than one identical fermion can occupy a
single state due to the Pauli exclusion principle. 
However, a mixed gas of fermions and bosons provides an effective means of cooling
single-component fermions by thermal collisions with evaporatively cooled 
bosons \cite{Fermi-1,Fermi-2,Fermi-3,Fermi-4}, providing an avenue for 
investigating many-body quantum effects in degenerate Fermi gases.
Another development is the use of Fesh\-bach resonance, in which the
energy of a bound state of two colliding atoms
is magnetic field tuned to vary the scattering strength from $-\infty$ to $\infty$,
allowing the investigation of the crossover from BCS superfluidity to 
Bose-Einstein condensation \cite{BEC-F1,BEC-F2,BEC-F3}.
These achievements in realizing quantum Fermi gases of ultracold atoms
may also provide insights into other areas of physics, such as
ultracold superstrings \cite{superstring}.

Particular theoretical and experimental interest has been paid to the 
Fermi gas confined in 1D geometry \cite{Fermi-1D1,Fermi-1D2,BEC-BCS1,BEC-BCS2,BEC-BCS3}.  
Recent attention has turned to the 1D model of mixed bosons and 
polarized fermions \cite{mix-1,mix-2,mix-3,mix-4}, revealing quantum
effects such as interaction-driven phase separation \cite{phase-s1,mix-3,mix-4}, bright solitons in degenerate Bose-Fermi mixtures \cite{kar04} and 
Luttinger liquid behaviour \cite{mix-1,mix-2}. 
In this model, only $s$-wave scattering for boson-boson and boson-fermion interactions
is considered, with the boson-fermion collisions minimizing the Pauli-blocking effect in 
momentum space.

The theoretical studies of the mixed Bose-Fermi gas have focused on the case of 
bosons and single-component fermions \cite{mix-1,mix-2,mix-3,mix-4}. 
Our aim here is to investigate the $1$D model of a Bose-Fermi mixture on a 
line of length $L$ with periodic boundary conditions, with Hamiltonian
$H=-\sum_{i=1}^N \frac{\hbar^2}{2 m_i} \frac{\partial^2}{\partial x_i^2} + \sum_{i<j}g_{i,j} \delta(x_i-x_j)$,
where among the $N$ particles there are $N_f=N_\down+N_\up$ fermions and $N_b$ bosons. 
The pairwise $\delta$-interaction has strength $g_{i,j}$. 
The crucial observation from a theoretical point of view is that if all particles have equal masses 
and if the interaction strength between all particles is the same, 
the above model is exactly solvable by Bethe Ansatz (BA) \cite{L-Y}. 
Although this restricts the applicability of the model, experiments with isotopes of atoms 
are expected to meet the BA conditions \cite{mix-4}. 
Then $g_{i,j}=2 \hbar^2/(m\,a)$, where $c=2/a$ is the inverse 1D scattering length of the confined particles. 
For convenience of notation, the energy is measured in units of $\hbar^2/(2 m)$, such that
\be
H=-\sum_{i=1}^N \frac{\partial^2}{\partial x_i^2} + \sum_{i<j}2 c \delta(x_i-x_j)\label{ham}.
\ee

The model contains two special limiting cases: 
(i) the $1$D interacting Fermi gas with arbitrary polarization 
\cite{Yang-Gaudin,Takahashi,Wadati,BBGN}, and
(ii) the $1$D interacting Bose gas \cite{LL63}.
The mixed model was recently discussed for the case of single component (fully polarized) fermions \cite{mix-4}.
Here we consider the more general case of two-component fermions with arbitrary polarization. 
We present analytical and numerical results for the ground state energy, 
quasimomentum density profile, and the spin and charge velocities.
In the weak coupling limit, these quantities reveal the typical signatures of pure quantum gases 
with an additional weak interaction due to the mixture. 
In the other extreme, that is for strong repulsive interactions, the ground state properties 
resemble those of a single-component non-interacting Fermi gas.

The BA equations (BE), determining the quantum numbers of the $N$-particle system, are given by  \cite{L-Y}
\be
\te^{{\rm i}\, p_\ell L}&=& \prod_{j}\! \frac{p_\ell-\L_j+\rmi c'}{p_\ell-\L_j- \rmi c'}\label{ba1}\\
\!\!\!\prod_{\ell}\frac{\L_k-p_\ell-\rmi c'}{\L_k-p_\ell+\rmi c'}\!\!\!&=&\!\!\!-
\prod_{j,m}\! \frac{\L_k-\L_j-\rmi c}{\L_k-\L_j+\rmi c}\frac{\L_k-A_m+\rmi c'}{\L_k-A_m-\rmi c'}\label{ba2}\\
\!\!\!1\!\!\!&=&\!\!\!\prod_{j}\!\frac{A_n-\L_j-\rmi c'}{A_n-\L_j+\rmi c'}\label{ba3},
\ee
where $c'=c/2$ and $j,k=1,\ldots,N_b+N_\up$, $\ell=1,\ldots,N$, $m,n=1,\ldots,N_b$. 
Without loss of generality, we take $N_\up<N_\down$. 
Here the set of $N+N_\up+2 N_b$ many quantum numbers is divided into three subsets: 
$\{ p_\ell\}_{1,\ldots,N}$, $\{ \Lambda_k\}_{1,\ldots,N_\up+N_b}$, $\{A_n\}_{1,\ldots,N_b}$. 
It turns out that the energy eigenvalues $E$ are given by the members of the first set alone, namely $E=\sum_{\ell=1}^N p_\ell^2$.

We begin with a finite system, for which the Bethe roots are found analytically in the 
weak coupling limit, thereby yielding the ground state energy. 
The second step is to carry out the thermodynamic limit (TL). 
In this limit, the Bethe roots are distributed 
smoothly over a certain interval of the real axis, giving rise to continuous densities.  
Integral equations for these densities have been obtained previously \cite{L-Y}. 
On the one hand, these equations allow a comparison between the analytical results for weak coupling 
and finite systems with the weak coupling expansion in the TL. 
On the other hand, generalized velocities can be calculated within this framework. 
In the weak coupling limit, these
reduce to the spin and charge velocities of a pure Fermi gas and 
the charge velocity of a pure bosonic system.

In considering the ground state for weak interaction, it is convenient to distinguish between 
unpaired $p_j^{(\rm u)}$ ($j=1,\ldots,N_\down-N_\up$), 
paired $p_j^{(\rm p)}$ ($j=1,\ldots,N_\up-1,N_\up+2,\ldots,2N_\up$) 
and bosonic $p_j^{(\rm b)}$ ($j=1,\ldots,N_b+2$) quasimomenta \cite{note1}.
Expanding Eqs.~\refeq{ba1}-\refeq{ba3} to $\mathcal{O}(c)$, 
one obtains $p_j^{(\rm u)}=\pi(-N_\down-1+2j)/L+\delta_j^{(\rm u)}$,
$j=1,\ldots,(N_\down-N_\up)/2$; 
$p_j^{(\rm u)}=\pi(2N_\up-N_\down-1+2j)/L+\delta_j^{(\rm u)},\,
j=(N_\down-N_\up)/2+1,\ldots,N_\down-N_\up$ and 
$p_j^{(\rm p)}=\pi (-1-N_\up+2j_+)+ \delta^{(\rm  p)}_{j_+}\pm \sqrt{c/L}$.
Here $j_+=j$ if $j$ odd and $j_+=j-1$ if $j$ even. 
The deviations $\delta_j$ from $p_j^{(\rm u,\rm p)}$ are linear in $c$, with 
\be
\!\!\!\delta_j^{(\rm u)}\!\!\!&=&\!\!\!\frac{c}{L}\!\l[\!\sum_k 
\frac{1}{p_{j,0}^{(\rm u)}-p_{k,0}^{(\rm p)}}+\frac{N_b+1}{p_{j,0}^{(\rm u)}}\!\r],
\label{dju}\\
\!\!\!\delta_j^{(\rm p)}\!\!\!&=&\!\!\!\frac{c}{L}\!\l[\!\sum_{k\neq j} \frac{1}{p_{j,0}^{(\rm p)}-p_{k,0}^{(\rm p)}}\!+\!\sum_{k} \frac{1/2}{p_{j,0}^{(\rm p)}-p_{k,0}^{(\rm u)}}\!+\!\frac{N_b+1}{p_{j,0}^{(\rm p)}}\!\r].\,\,
\label{djp}
\ee
Here $p_{j,0}^{(\rm p,\rm u)}$ denotes the quasimomenta in the free particle limit as given above.  
The sums in \refeq{dju}, \refeq{djp} over paired momenta $p_{k,0}^{(\rm p)}$ count each pair only once.  
The bosonic momenta $p_j^{(\rm b)}$ collapse at the origin for $c=0$. 
Their $c$-dependence is given by
\be
p_j^{(\rm b)}= \frac{2 c}{L} \sum_{k\neq j} \frac{1}{p_{j}^{(\rm b)}-p_k^{(\rm b)}}\label{pjb}.
\ee
According to \refeq{pjb}, the $p_j^{(\rm b)}$ are the roots of Hermite polynomials of degree 
$N_b+2$ \cite{gau71,BGM}.

{}From the above equations we obtain the ground state energy, $E=E^{(0)}+E^{(1)}$, 
to leading order in $c$. 
The energy of the free particles, $E^{0}$, is given by the corresponding expression for a 
free Fermi gas with $N_f$ fermions \cite{BBGN}. 
The first correction in $c$ is $E^{(1)}=E^{(1)}_b+E^{(1)}_f+2\frac{c}{L}N_bN_f$, 
where $E^{(1)}_b$ ($E^{(1)}_f$) is the linear order in $c$ for a pure Bose (Fermi) gas of 
$N_b$ bosons \cite{BGM} ($N_f$ fermions \cite{BBGN}).
The last term is due to the interactions between bosons and fermions. 
Thus
\be
E&=& \l(\frac{2\pi}{L}\r)^2\l(\frac16N_\up(N_\up^2-1)+\r.\nn\\
& & \l.\frac{1}{12}(N_\down-N_\up)(-1+N_\up^2+N_f^2-N_fN_\up)\r)\nn\\
& & +\frac{2c}{L}\l(N_\down N_\up +\frac12 N_b(N_b-1) +N_bN_f\r).\label{en}
\ee
This expression is valid for both weak repulsive and attractive interaction. 
Especially, the TL is well defined in the weakly attractive case, 
as opposed to the strongly attractive case \cite{Takahashi}. 
Let us now carry out the TL, i.e., $N_\alpha,L\to\infty$ where the densities 
$n_\alpha=N_\alpha/L$ are held constant, with $\alpha=\down,\up,b$. 
In this limit, Eqs. \refeq{dju} and \refeq{djp} give the distribution 
of quasimomenta per unit length,
\be
\rho_u(p)\!\!\!&=&\!\!\!\frac{1}{2\pi} +\frac{c}{2\pi^2}\l(\frac{A}{p^2-A^2}+\frac{B}{p^2}\r),\,A<|p|<C\label{nu}\\
\rho_p(p)\!\!\!&=&\!\!\!\frac{1}{\pi}-\frac{c}{2\pi^2}\l(\frac{A}{A^2-p^2}+\frac{C}{C^2-p^2}-\frac{2 B}{p^2}\r),\nn\\
& &\,2\sqrt{cB/\pi}<|p|<C\label{np}\\ 
\rho_b(p)\!\!\!&=&\!\!\!\frac{1}{2\pi c} \l(4 c B/\pi-p^2\r)^{1/2},\,|p|<2\sqrt{cB/\pi}\label{nb},
\ee
where $A=\pi N_\up/L$, $B=\pi N_b/L$, $C=\pi(N_\down-N_\up)/L$.

\begin{center}
\begin{figure}[t]
\epsfig{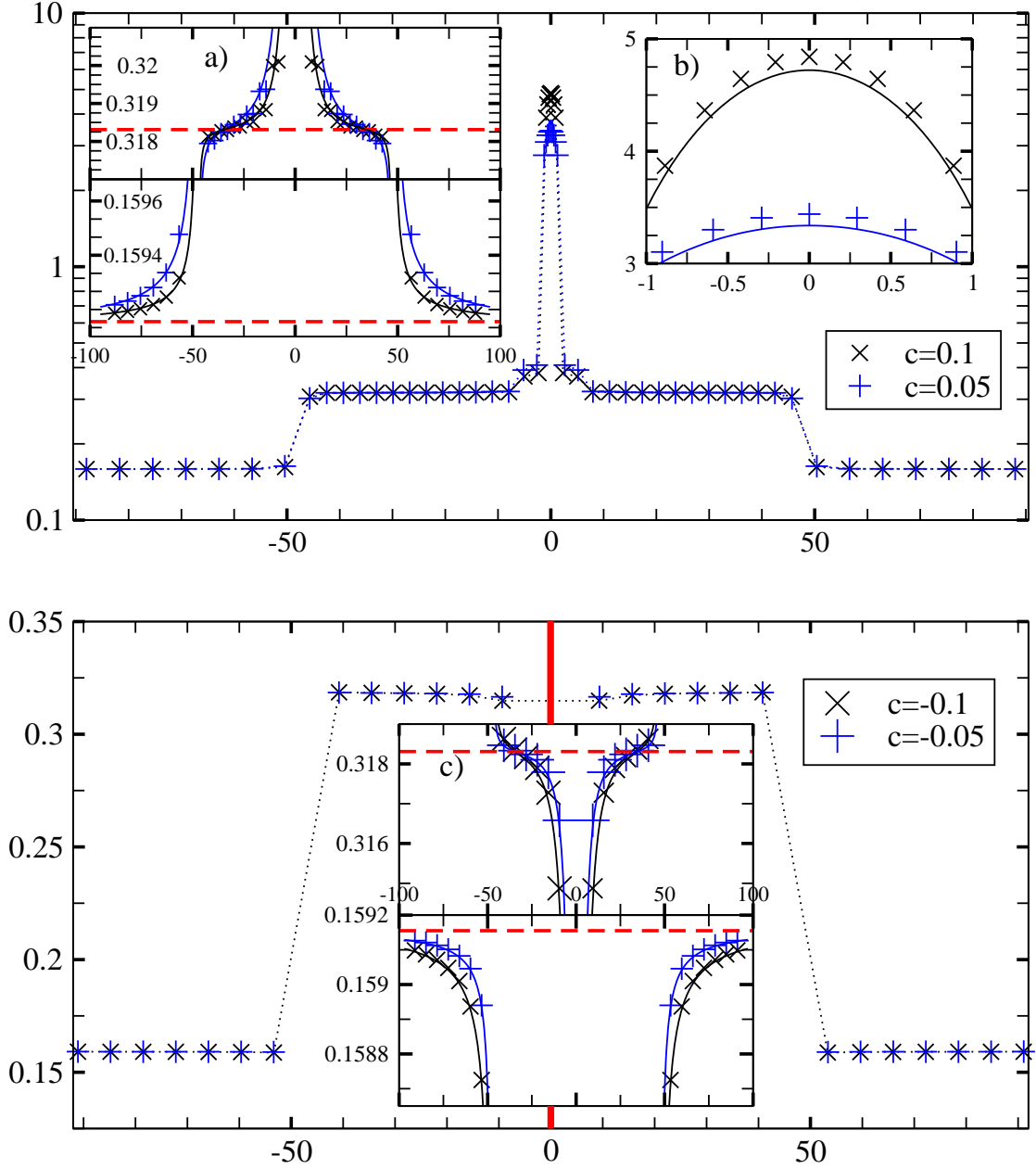}
\caption{The quasimomentum distribution functions for 
a finite system (crosses) ($N_\up=15, N_\down=31, N_b=9$) and in the TL (full lines in the insets) 
for (top) weakly repulsive and (bottom) weakly attractive interactions. 
The insets a),c) show the distributions near the
noninteracting cases (dashed lines) $1/2\pi$ for unpaired fermions and $1/\pi$ for paired fermions. The inset b) depicts the bosonic quasimomenta for $c>0$; for $c<0$, their real part is zero (thick line).}
\label{Fig1}
\end{figure}
\end{center}

The quasimomentum distribution function of the bosonic momenta \refeq{nb} for $c>0$ is given by 
the semi-circular law behaviour as for a pure bosonic system \cite{LL63,gau71,BGM}. 
For $c<0$, the $p_j^{(\rm b)}$ in \refeq{pjb} are imaginary, so that \refeq{nb} with imaginary $k$ yields the dark-soliton like distributions of the bosonic quasimomenta on the imaginary axis. 
These encode the binding energy of the bosons, a quantity potentially accessible by experiments. 
In the distribution of the fermionic quasimomenta, divergences occur near the cutoffs. 
The quasimomentum distribution functions calculated from the BE \refeq{ba1}-\refeq{ba3} are 
compared with the approximations \refeq{nu}-\refeq{nb} in Fig. \ref{Fig1}.

\begin{center}
\begin{figure}[t]
\epsfig{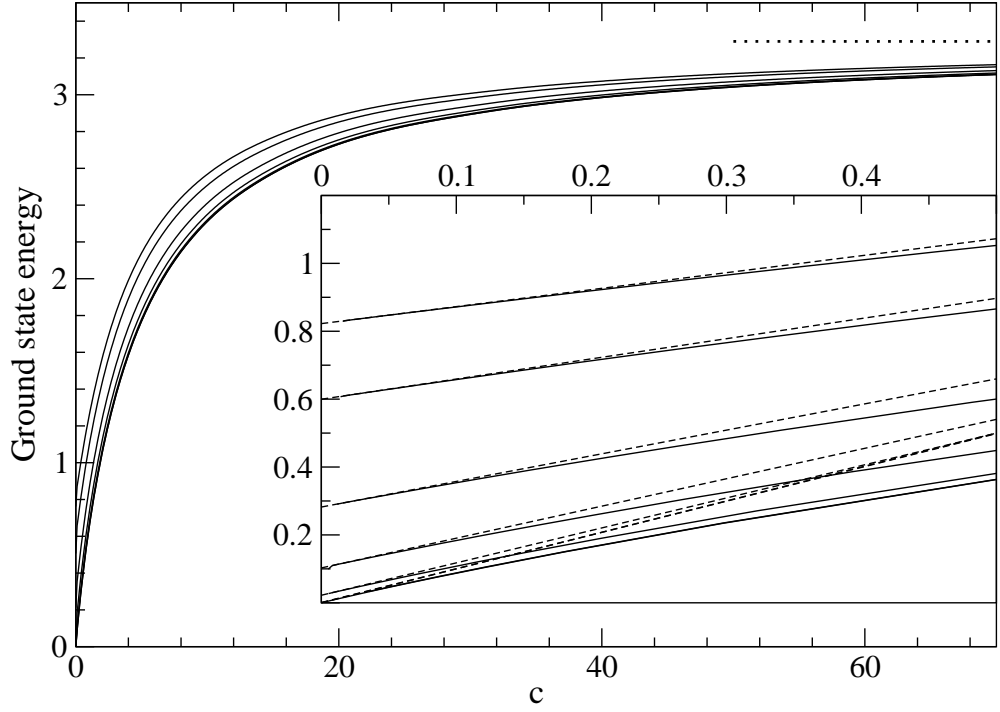}
\caption{Ground state energy per unit length $E/L$ in the TL as a function of the interaction strength
for densities $n_\up=n_\down$, $n=n_b+n_f=1$, with $n_b=1,0.9,0.7,\ldots,0.1,0$ from bottom to top. 
The dotted line is the asymptotic result $E/L\to \pi^2/3 n^3$ for $c\to \infty$, 
which is equal to the energy of free fermions with density $n$.
The inset compares the analytic \refeq{en} (straight dashed lines) with numerical TL-results (solid lines) 
for weak repulsive interaction.}
\label{Fig2}
\end{figure}
\end{center}

In order to study the effect of arbitrary interaction in the TL, we use the linear integral equations 
derived by Lai and Yang \cite{L-Y} which determine the densities of the quasimomenta.
In the following, we restrict ourselves to two limiting cases, namely $N_\up=N_\down$ and $N_\up=0$. 
In the latter case, the fermions do not interact among themselves due to the Pauli principle, so that the interaction potential in \refeq{ham} is only effective 
between fermions and bosons. 
The ground-state energy for this case has been calculated recently \cite{mix-4}. 
In the other limiting case, $N_\up=N_\down$, all the fermions interact with each other and with the bosons. 
Fig.~\ref{Fig2} shows the ground-state energy per unit length $E/L$ for different densities $n_b$ and $n_f$ 
obtained from numerical solution of the integral equations.
Also shown is a comparison between the analytic result \refeq{en} and the TL. 

Up to this point, we have focussed on the density $\rho=\rho_u+\rho_p+\rho_b$ of the quasimomenta $p_\ell^{(\rm u,\rm p,\rm b)}$. 
In an analogous fashion, one may introduce densities $\sigma$, $\tau$ of the roots $\L_j, A_m$. 
Using the dressed energy formalism \cite{Takahashi}, one can calculate the corresponding dressed 
energies $\epsilon, \phi, \psi$, which are the energies necessary to add or remove a root to or from 
the seas of $p_\ell, \Lambda_j, A_m$.
The dressed energies give rise to generalized velocities, which determine the asymptotics of correlation functions \cite{frahm}.  
In the case of pure bosons, there is only one species of BE numbers, associated with one dressed energy function, yielding the charge (or sound) velocity $v_c^{(b)}$ \cite{LL63,Takahashi}. 
For pure fermions, the two sets of BE numbers are linked with the charge and 
spin velocities $v_{c,s}^{(f)}$. 
As has been proven by Haldane \cite{hal81a}, these velocities coincide with those calculated in 
the harmonic-fluid (or bosonization) approach \cite{hal81b}.

\begin{center}
\begin{figure}[t]
\epsfig{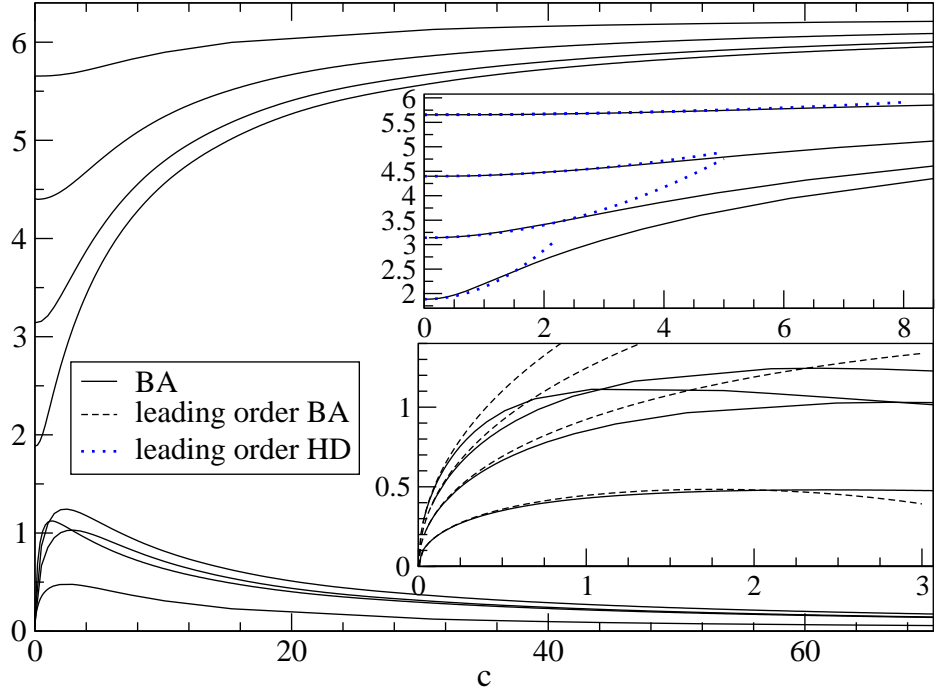}
\caption{Velocities $\tilde v_c^{(f)}$ (upper set of curves) and $\tilde v_c^{(b)}$ (lower set of curves) 
for polarized fermions with $n_f=0.9,0.7,0.5,0.3$ (from top to bottom) interacting with bosons $n_b=0.1,0.3,0.5,0.7$ (from bottom to top). 
The insets show the weakly repulsive limit, together with analytical results from BA and HD.}
\label{Fig3}
\end{figure}
\end{center}

The situation is less clear for the mixture of bosons and fermions. 
In the BA approach, this corresponds to two seas of BE numbers $p_\ell,\Lambda_j$, 
and correspondingly two dressed energies and two velocities, which we call $\tilde v_c^{(b,f)}$. 
For $c=0$ we have $\tilde v_c^{(f)}=2\pi n_f$ and $\tilde v_c^{(b)}=0$.
The dependence of $\tilde v_c^{(b,f)}$ on $c$ for different values of $n_f, n_b$ is shown in Fig.~\ref{Fig3}. 
The lowest $c$-order of $\tilde v_c^{(b)}$ is obtained from the Bethe-Ansatz as $\tilde v_c^{(b)}=2 n_b\l[c/n_b-(1/2\pi)(c/n_b)^{3/2}\r]^{1/2}$, 
which is the velocity of a non-interacting Bose gas for small $c$ \cite{LL63}.

\begin{center}
\begin{figure}[t]
\epsfig{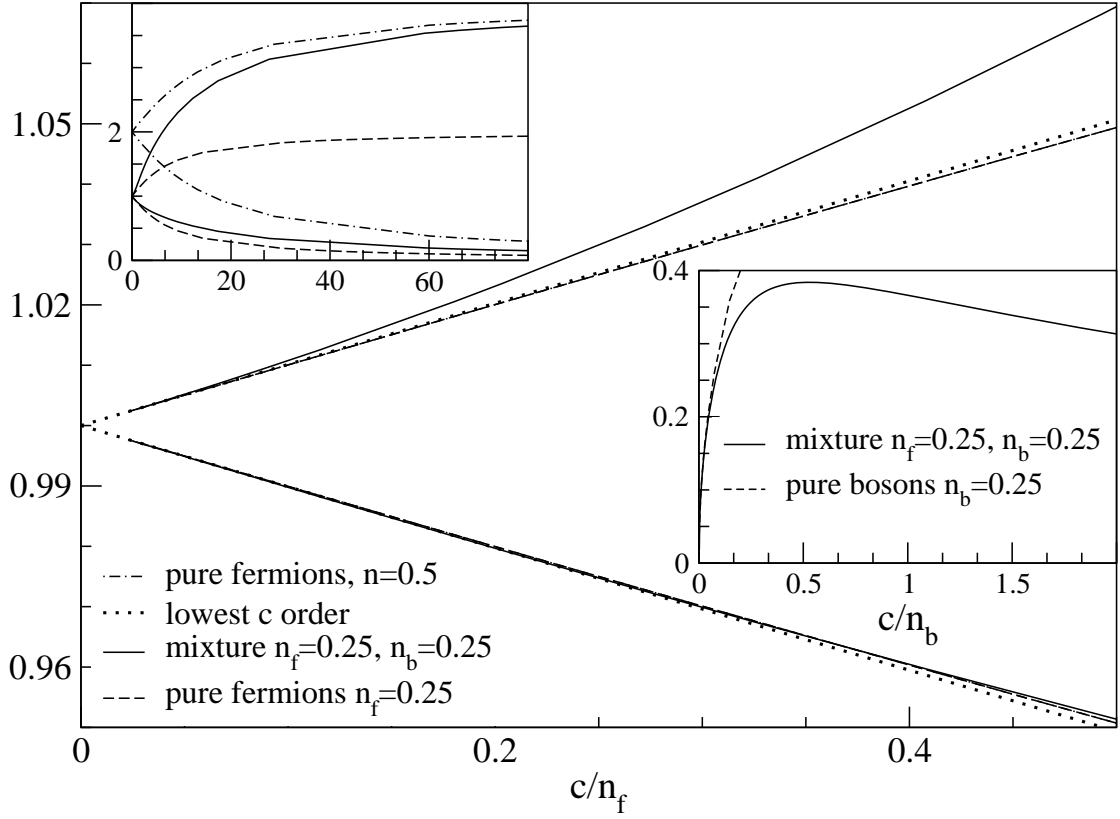}
\caption{Velocities $\tilde v_{c}^{(f)}/(\pi n_f)$ (upper curve) and $\tilde v_s^{(f)}/(\pi n_f)$ (lower curve) for 
a mixed Bose-Fermi gas with $n_f=n_b=0.25$ plotted against $c/n_f$. 
The upper inset shows the whole $c$-range. 
The second inset shows $\tilde v_c^{(b)}$ for weak couplings.}
\label{Fig4}
\end{figure}
\end{center}

We now compare our results to those of the hydrodynamic (HD) approach \cite{mix-1}. 
In this framework, the pure gases correspond to harmonic oscillators, such that the coupling 
between them leads to new normal modes $v_\pm$, with $v_\pm(c=0)=\tilde v_c^{(f,b)}(c=0)$. 
In the weak interaction limit, $v_-=2\sqrt{c n_b}$, $v_{+}=2\pi n_f+\frac{n_bc^2}{n_f^2 \pi^3}$. 
For the sake of comparison, the latter result is also shown in Fig.~\ref{Fig4}. 
In the strongly repulsive limit,
a demixing instability is predicted from the HD \cite{mix-1} and mean-field approaches \cite{mix-3}. 
We do not observe any instability for repulsive interaction for the integrable model, 
in agreement with \cite{L-Y,mix-4}.  
The reason for the discrepancy is that to our present understanding, the HD approach for the mixture, 
especially the calculation of normal modes, is a low-energy theory. 
That is, it is expected to yield reliable results for small interaction strengths $|c|\ll 1$. 
%
Investigation of $\tilde v_c^{(f,b)}$ within the BA approach in the strongly repulsive limit yields
\be
\tilde v_c^{(f)}&=&2\pi n \l[1-\frac{4 n}{\pi c} \l[ \frac{\pi n_b}{n}+\sin\l(\frac{\pi n_b}{n} \r) \r] \r],\nn\\
\tilde v_c^{(b)}&=&\frac{4 \pi^2 n^2}{3 c} \,\sin\l(\frac{\pi n_b}{n}\r)\nn,
\ee 
where $n=n_f+n_b$.

It remains a very interesting question for future research if the normal modes of the field theory coincide with the generalized velocities from the BA 
for mixed Bose-Fermi systems. 
Furthermore, it should be carefully investigated whether or not the HD approach is applicable in the 
strongly interacting regime.

The generalized velocities can also be computed from the BA for interacting fermions ($n_\up=n_\down$). 
In this case, the three dressed energies $\epsilon,\phi,\psi$ give rise to three velocities, 
which we call $\tilde v_{c,s}^{(f)}$ and $\tilde v_{c}^{(b)}$. 
As shown in Fig.~\ref{Fig4}, 
we observe that in the weak interaction limit, $\tilde v_{c,s}^{(f)}=\pi n_f(1\pm c/(n_f \pi^2))$, 
which corresponds to the charge and spin velocities of a pure fermionic system \cite{BEC-BCS1}. 
In the strong interaction limit, $\tilde v_{c}^{(b)}=\tilde v_{s}^{(f)}=0$, whereas $\tilde v_{c}^{(f)}=2\pi n$. 
This again indicates the fermionic nature of the system in the strongly interacting limit.

In conclusion, we have investigated ground state properties of the 1D interacting Bose-Fermi 
model from its exact BA solution. 
We obtained results for the distribution of quasimomenta and the ground state energy both for 
weak attractive and repulsive interactions. 
We computed the generalized velocities and compared them to the normal modes obtained 
from the HD approach. 
In contrast with other approaches \cite{mix-1,mix-3,mori},
we do not observe any instability or demixing in the system.
%

{\em Acknowledgements.}  This work has been supported by the Australian Research
Council and the German Science Foundation under grant number BO/2538.

\begin{thebibliography}{99}
\bibitem{note1} Since the substitution of two fermions into $N_b$ bosons is equivalent to 
increasing $N_b$ to $N_b+2$, there are $N_b+2$ momenta $p_j^{(\rm b)}$.
\end{thebibliography}
\end{document}